\documentclass{rspublic}
\usepackage{graphicx}
\usepackage{epsfig}
\usepackage{psfrag}
\usepackage{bm}
\usepackage{amsbsy}
\usepackage{amssymb} 
\usepackage{lineno}

\def\dd{\mbox{d}}

\def\dd{\mbox{d}}

\newcommand{\pt}[2]{\frac{\partial{#1}}{\partial{#2}}}

\begin{document}

\title[Reconstruction of potential energy profiles]
{Reconstruction of potential energy profiles
from multiple rupture time distributions}

\author[P.-W. Fok and T. Chou]{Pak-Wing Fok$^{\dag}$ and Tom Chou}
\affiliation{Dept. of Biomathematics and Dept. of Mathematics, UCLA, Los
Angeles, CA 90095-1766}
\footnotetext[1]{Present Address: Department of Mathematical Sciences, University of Delaware, 
Newark, DE 19716-2553}
\label{firstpage}

\maketitle


\begin{abstract}{inverse problem, first-passage time, bond rupture}
We explore the mathematical and numerical aspects of reconstructing a
potential energy profile of a molecular bond from its rupture time
distribution.
While reliable reconstruction of gross attributes, such as
the height and the width of an energy barrier, can be easily extracted
from a single first passage time (FPT) distribution, the
reconstruction of finer structure is ill-conditioned.  More
careful analysis shows the existence of optimal bond potential
amplitudes (represented by an effective Peclet number) and initial
bond configurations that yield the most efficient numerical
reconstruction of simple potentials. Furthermore, we show that
reconstruction of more complex potentials containing multiple minima
can be achieved by simultaneously using two or more measured FPT
distributions, obtained under different physical conditions. For
example, by changing the effective potential energy surface by known
amounts, additional measured FPT distributions improve the
reconstruction. We demonstrate the possibility of reconstructing
potentials with multiple minima, motivate heuristic rules-of-thumb for
optimizing the reconstruction, and discuss further applications and
extensions.
\end{abstract}

\section{Introduction}

In many applications, one wishes to infer properties of a material or
a process in an interior region of a sample not readily accessible to
experimental probes.  Examples of such inverse problems involving
boundary data include radiological imaging, where radiation passing
through tissues is detected outside the sample, electrical impedance
tomography, where potentials are measured on the exterior of a body,
and seismology, where reflected waves are measured at the earth's
surface.  Such problems are often
ill-conditioned: there may be several different interior structures
that yield nearly the same measured boundary data.

\begin{figure}[htbp]
\begin{center}
\includegraphics[width = 4in]{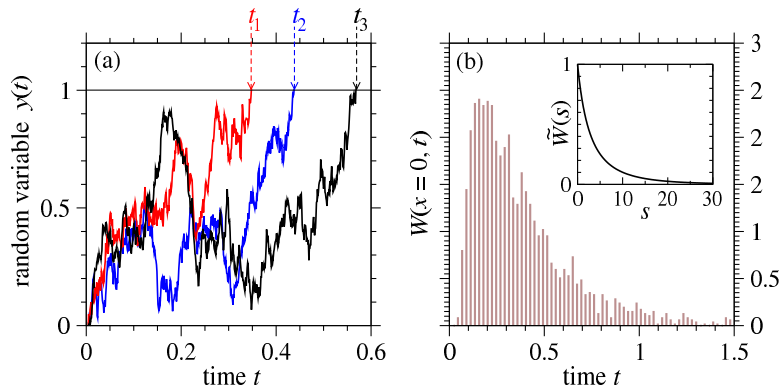}
\end{center}
\caption{(a) Three simulated realizations of a representative random
walk and their first passage times $t_i$. The random variable $y(t)$
could represent the transmembrane voltage of a neuron or the bond
coordinate of an unfolding macromolecule.  (b) Histogram of the first
passage times of a stochastic process started at position $x=y(0)=0$,
$W(x=0,t)$, obtained from 2000 realizations of the process shown in
(a). The inset shows $\tilde{W}(s) =
\int_{0}^{\infty}W(t)e^{-st}\dd t$, which is used extensively in
this paper.  An arbitrary potential was used to generate the data.}
\label{FIG1}
\end{figure}

One type of ``boundary'' data that often arises in stochastic models
is a first passage time distribution (FPTD), describing the
probability of a random variable first reaching a particular value
within a certain time window. Here, the boundary data is the
probability flux out of the domain. Figure \ref{FIG1}(a) shows
individual trajectories of a one-dimensional stochastic process and
their corresponding first passage times. The FPTD is shown in figure
\ref{FIG1}(b) along with its Laplace transform in the inset. These
types of first passage problems arise in many biophysical
contexts. For example, the voltage across a nerve cell membrane
fluctuates due to noisy inputs from other neurons, and can be
described by a biased random walk determined by a constitutive
voltage-current relationship intrinsic to the cell (Tuckwell
\textit{et al.} 2003) When the fluctuating voltage exceeds a
threshold, the potential rapidly spikes before resetting.  The
interspike times define the first passage times of the fluctuating
voltage from which one might wish to reconstruct the neuron's inherent
current-voltage relationship.

Stochastic inverse problems are typically ill-posed: there may be
several different interior structures that could yield identical or
nearly identical measured boundary data. Nonetheless, for many
physical systems, reconstruction of constitutive relations from
measured data can be cast in Sturm-Liouville form with an unknown
spatially dependent coefficient (Levitan 1987, McLaughlin 1986) Given
the eigenvalues of the problem and assuming a symmetric coefficient
function, its full reconstruction is unique (Borg 1946). However, one
eigenvalue spectrum is insufficient to determine a general
nonsymmetric coefficient (Borg 1946).
%
%
For stochastic problems, the spectrum of the corresponding
Sturm-Liouville problem cannot be readily extracted from data and
algorithms developed specifically for reconstruction through the
eigenvalues (Brown \textit{et al.} 2003, Rundell \& Sacks 1992$a$,
Rundell \& Sacks 1992$b$) are of limited use in our stochastic
problem.  This motivates the development of new algorithms and
techniques that deal directly with the boundary data.

In this paper, we investigate a stochastic inverse problem in the
context of another system commonly encountered in biophysics:
macromolecule unfolding and molecular adhesion bond rupturing.
Macromolecular bond displacements are often described by a single bond
coordinate, represented by a fluctuating Brownian ``particle'' in a
one-dimensional energy landscape. The metastable bond is considered to
be broken the instant the bond coordinate reaches a critical
extension. This problem is of great interest in single-molecule
biophysics, particularly in the context of dynamic force spectroscopy
(DFS) (Evans \textit{et al.} 1995). In DFS, a pulling force protocol
is applied to the bond and the force at the instant of rupture is
recorded. The mean rupture force by itself would give very little
information about the molecular potential (Schlierf \& Rief 2006)
since many different potentials would yield the same mean rupture
force.  How much of the bond potential can be recovered from the
measured rupture force {\it distribution}?  All of the recent
theoretical treatments of this problem have either analysed the
forward problem (Heymann \& Gr\"{u}bmuller 2000), used physical
approximations to derive simple force-dependent and time-dependent
dissociation rates (Bell 1978, Walton \textit{et al.} 2008), and/or
considered simple 2-3 parameter single minimum potentials (Hummer \&
Szabo 2003, Dudko \textit{et al.} 2008, Freund 2009). The rate of
force increase as a function of displacement (the rupture stiffness)
has also been incorporated into a procedure to fit basic parameters of
simple potentials (Fuhrmann \textit{et al.} 2008). However, by
imposing such simple two or three parameter forms for the
reconstructed potential, one loses details such as multiple minima.

Here, we approach the inverse problem by allowing a wider class of
potentials, including those with multiple minima.  Within a class of
potentials, we numerically determine the ones that best fit the entire
measured FPTD. Although the difficulty of extracting eigenvalues from
FPTD data is avoided, the inverse problem remains intrinsically
ill-conditioned and it is not surprising that almost all studies have
focused on reconstructing only two or three attributes of the
stochastic process, typically, the energy barrier height and width. In
\S2, we formulate the problem through the backward equation of a
Brownian process with a potential energy-derived drift. In \S3, we
decompose the drift function into basis functions and develop an
iterative optimization procedure to find the coefficients of these
basis functions.  In \S4, we show that using a single FPTD restricts
the type of potentials we can reconstruct. We also show how the
inverse problem can be optimized by tuning the amplitude of the
unknown potential and the initial bond displacement. Another key
finding is that multiple FPTDs greatly facilitate the reconstruction,
allowing us to accurately determine potentials with multiple
minima. We propose experimental protocols that can be used to generate
these additional FPTDs.  Finally in \S5, we discuss limitations of our
method and possible refinements.

\section{Stochastic Theory and the Inverse Problem}

The general problem of stochastic bond rupturing is geometrically
complex, particularly when considering deformations associated with
large macromolecules carrying many degrees of freedom. Although in
principle these systems can be modeled by stochastic processes in
higher dimensions, for simplicity, and we restrict our mathematical
analysis to a one-dimensional Brownian motion described by a
diffusivity $D(x)$ and a convective drift $-D(x)(k_{B}T)^{-1}(\dd
\Phi(x)/\dd x)$ proportional to the force derived from a
time-independent molecular bond potential $\Phi(x)$, and to the
mobility $D(x)(k_{B}T)^{-1}$.  Although we restrict ourselves to
time-independent potentials, corresponding to static forces, the
rupture force distribution can be transformed into a first rupture
time distribution (FPTD) in the quasi-adiabatic limit (Dudko
\textit{et al.} 2008).  The continuous Brownian process can be
described by the probability $P(y,t\vert x)\dd y$ for the bond
coordinate to be between positions $y$ and $y+\dd y$ at time $t$,
given that it started at position $x$ at initial time $t=0$. This
probability density obeys the backward equation (Gardiner 2004)

\begin{equation}
{\partial P(y,t\vert x)\over \partial t} + {D(x)\over k_{B}T}{\dd
\Phi\over \dd x}\left({\partial P(y,t\vert x) \over \partial
x}\right) = D(x){\partial^{2}P(y,t\vert x)\over \partial x^{2}}.
\label{BACKWARD1}
\end{equation}
Since the bond is irreversibly ruptured when stretched past a known
position $y=L$, we impose the absorbing boundary condition
$P(y=L,t\vert x) =0$. The bond survival probability at time $t$, given
that it started initially at position $x$ is found from integrating
the probability density over all the final coordinates that an
unruptured bond can take, {\it i.e.}, $S(x,t) = \int_{0}^{L}P(y,t\vert
x)\dd y$. From $S(x,t)$, we define the FPTD $w(x,t) =
-\partial_{t}S(x,t)$, which obeys

%
\begin{equation}
{\partial w(x,t)\over \partial t} + {D(x)\over k_{B}T}{d \Phi(x)\over d x}
{\partial w(x,t)\over \partial
x} = D(x) {\partial^{2}w(x,t) \over \partial x^{2}},
\label{PDE1}
\end{equation}
subject to initial condition $w(x,0) = 0$, and boundary conditions
$\partial_{x} w(x,t)\vert_{x=0} =0$ and $w(x=L,t) = \delta(t)$.

\begin{figure}
\begin{center}
\includegraphics[width = 2.4in]{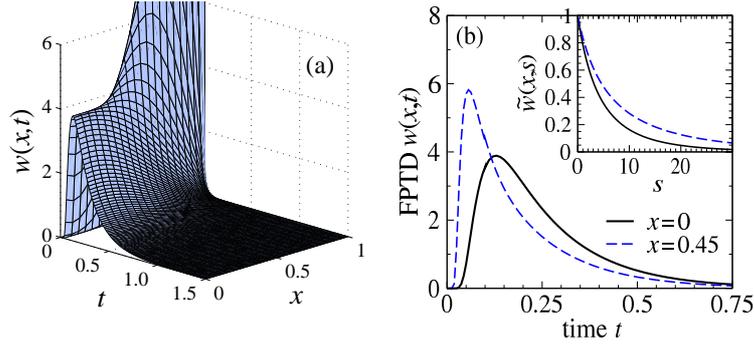}\hspace{-7mm}
\includegraphics[width = 1.85in]{Fig2b.eps}
\end{center}
\caption{(a) The solution to the forward equation \ref{PDE1} for the
first passage time distribution (FPTD) as a function of starting
position $x$ with $D=1$ , $L=1$ and $U(x) = 4 - 10x+6x^2$. The delta
function at $t=0$ when $x=1$ was approximated by a sufficiently narrow
gaussian centered at $t=0$.  (b) For each starting position $x$,
slices through the surface define the FPTD $w(x,t)$.  Two slices
corresponding to the starting positions $x=0$ (solid) and $x=0.45$
(dashed) are shown. Inset: The Laplace transform of the same two
slices.}
\label{fig:surface}
\end{figure}

In the forward problem, $\Phi(x)$ and $D(x)$ are given and one solves
equation (\ref{PDE1}) to find the function $w(x,t)$, as shown in
figure \ref{fig:surface}.  Each fixed-$x$ slice of the surface
$w(x,t)$ represents the FPTD for a particle that started the random
walk at position $x$.

In the inverse problem, the functions $\Phi(x)$ and $D(x)$ are unknown
and need to be determined from an experimentally measured or simulated
FPTD and a known starting position $x$. In general, the unique
pointwise reconstruction of both $D(x)$ and $\Phi(x)$ from FPT data is
impossible (Bal \& Chou 2003). At best, only half of either $D(x)$ or
$\Phi(x)$ can be uniquely determined from a single FPTD (Bal \& Chou
2003, Chen \textit{et al.} 1985).  It has been shown that if we assume
$D$ is a known constant, then $\Phi(x)$ is uniquely identifiable from
a single FPTD provided it is already known over a certain interval
within $(0,1]$ (Bal \& Chou 2003).

\section{Reconstruction Algorithm} 

Since in this problem, $\Phi(x)$ is not known on any requisite
interval, it is not even clear whether $\Phi(x)$ can be uniquely
reconstructed. Nevertheless in this section we express $\Phi(x)$ as a
superposition of (smooth) basis functions and attempt to reconstruct
the coefficients, deferring a rigorous analysis of our algorithm for a
future publication.  We find that representing $\Phi(x)$ using a
relatively small number of basis functions renders the problem
computationally tractable, yielding a unique solution in many cases.


Henceforth, we nondimensionalize the problem by measuring distance in
units of $L$, time in units of $L^{2}/D$, and the potential $\Phi(x)$
in units of the thermal energy $k_{B}T$.  Finally, to avoid
numerically representing the $\delta-$function in the boundary
condition $w(1,t)=\delta(t)$, we work with the Laplace transform
$\tilde{w}(x,s) = \int_{0}^{\infty}w(x,t)e^{-st}\dd t$, which obeys
the infinite set (for each $s\in \mathbb{R}_{\geq 0}$) of uncoupled
ODEs

\begin{equation}
{\partial^{2} \tilde{w}(x,s)\over \partial x^{2}} 
+ U(x){\partial \tilde{w}(x,s)\over \partial x} = s \tilde{w}(x,s), 
\label{ODE}
\end{equation}
subject to the Laplace-transformed boundary conditions
$\partial_{x}\tilde{w}(x,s)\vert_{x=0}= 0$ and 
$\tilde{w}(1,s) = 1$.

In equation (\ref{ODE}), we have defined the dimensionless drift $U(x)
\equiv -\dd \Phi(x)/\dd x$. The condition
$\partial_{x}\tilde{w}(x,s)\vert_{x=0}= 0$ represents a reflecting
boundary at $x=0$ and ensures a nonnegative bond coordinate. Equation
(\ref{ODE}) is a differential equation in the initial bond position
for the Laplace-transformed rupture time distribution
$\tilde{w}(x,s)$.
%

Mathematically, our objective is to reconstruct $U(x)$ and then
extract, modulo an irrelevant constant, the unknown potential
$\Phi(x)$.  However, to do so, we must choose a reduced representation
of $U(x)$ that renders, for a chosen value of $x$, and all $s \in
\mathbb{R}_{\geq 0}$, $\tilde{w}(x,s)$ as close as possible to the
measured or simulated Laplace-transformed FPTD $\tilde{W}(s)$. Since
it is known that approximating a function using a basis of monomials
leads to a very ill-conditioned problem (Keller 1975), we represent
the convection in terms of orthonormal polynomials:

\begin{equation}
U(x) = \sum_{i=0}^{n-1} a_{i} u_{i}(x),
\end{equation}
where $\{a_{i}\}\equiv {\bf a}$ is the vector of expansion
coefficients, and the first few orthonormal basis functions are
\begin{equation}
u_0(x) = 1, \,\, u_{1}(x) = \sqrt{3}(1-2x), \,\,
u_2(x) = \sqrt{5}(1-6x+6x^2), \ldots 
\label{UBASIS}
\end{equation}
Our method for reconstructing $U(x)$ consists of using a spectral
method (Trefethen 2000) to repeatedly solve the forward problem
equation (\ref{ODE}) to refine our estimate for the
potential.\footnote{More general potentials and drift functions that
diverge at $x=0$ lead to highly singular differential equations, but
can still be solved using spectral methods (Trefethen 2000).}

Starting with an initial guess for the drift (say, $U(x) = 0$, the
``null'' hypothesis), we solve equation (\ref{ODE}) for many positive
values of $s$ to obtain a numerical approximation for $\tilde{w}(x,s;
{\bf a})$.  We then compute the ``distance'' between the
$\tilde{w}(x,s; {\bf a})$ and the given data $\tilde{W}(s)$ using
the objective function
\footnote{If we had chosen to work in the time domain, a reasonable
objective function that measures the difference between measured and
computed FPTDs would be $\Pi({\bf a}) = \int_{0}^{\infty}\vert
w(x,t;{\bf a}) - W(t)\vert^{2} g(t) \dd t$.}

\begin{eqnarray}
\Pi({\bf a}) &=& \int_{0}^{\infty}  \vert \tilde{w}(x,s;{\bf a}) - 
\tilde{W}(s)\vert^2 g(s) \dd s,
\label{eqn:R1} 
\end{eqnarray}
where $g(s)$ is a function that weights FPT data differently for
different $s$.  By appropriately adjusting $U(x)$, implemented through
small changes in ${\bf a}$, $\Pi({\bf a})$ is decreased.  The
incremental adjustments in ${\bf a}$ are repeated until $\Pi({\bf a})$
is minimized. Although many different algorithms can be used to
minimize $\Pi({\bf a})$, we first consider $g(s)=1$ and choose a
safe-guarded Newton strategy that relies essentially on computing the
Hessian of $\Pi({\bf a})$. Details of the algorithm are described in
\ref{appA}.


\section{Results and Discussion}

We test our algorithm and discuss reconstructing the drift function
$U(x)$ from (i) a single, perfectly measured distribution of rupturing
times, and (ii) multiple perfectly measured distributions of rupturing
times, realized under different experimental conditions. We first
generate {\it perfect} ``data'' by solving the forward problem using a
hypothetical target potential function $\Phi^{*}(x)$ (and
corresponding $U^{*}(x) = \sum_{i=0}^{n-1}a_{i}^{*}u_{i}(x)$). After
generating the data $\tilde{W}(s) = \tilde{w}(x,s;{\bf a}^{*})$, we
pretend we did not know the coefficients ${\bf a}^{*}$, and try to
reconstruct them by minimizing $\Pi({\bf a})$ through successive
iterations $k$ of the numerical algorithm detailed in
\ref{appA}. Starting from an initial guess for ${\bf a}(k=0)={\bf 0}$,
we investigate if and how ${\bf a}(k)$ approaches ${\bf a}^{*}$, and
the number of coefficients $a_i$ that can be reliably reconstructed.
Using a single FPTD, we find that reconstruction of $\Phi^{*}(x)$ is
badly conditioned for $n \gtrsim 4$, but that using two or more
distinct FPTDs allows us to easily find 5 or more coefficients of
$\Phi^{*}(x)$ in many cases.


%
%
\subsection{Single measurement}

\begin{figure}[htbp]
\begin{center}
\includegraphics[width = 4.0in]{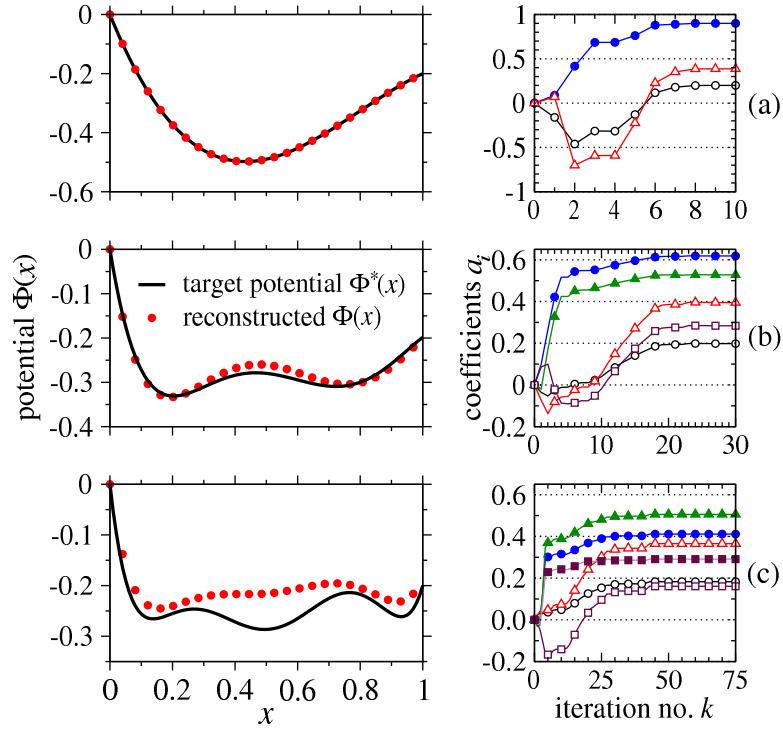}
\end{center}
\caption{Reconstruction of three, five and six parameter
potentials.
Row (a): reconstruction of a three parameter
potential corresponding to drift coefficients
$(a^{*}_0,a^{*}_1,a^{*}_2) = (1/5,9/10,\sqrt{3/20})$.
%
%
Row (b): attempted reconstruction of a five parameter double-well
potential 
with coefficients
$\mathbf{a}^* = (1/5,3/5,2/5,3/5,\sqrt{2}/5)$.  
%
Row (c): failed reconstruction of a six-parameter
potential defined by ${\bf a}^{*} = (1/5,
1/2, 2/5, 1/2, 1/5, \sqrt{13/50})$. In the second column, the
coefficient values $a_{0}, a_{1}, a_{2}, a_{3}, a_{4}, a_{5}$ at each
iteration are indicated by open circles, filled circles, open
triangles, filled triangles, open squares, and filled squares,
respectively.}
\label{fig:phi1}
\end{figure}

We first assume a target potential $\Phi^{*}(x) = -({1+7\sqrt{3}\over
5})x + {12\sqrt{3}\over 5}x^{2} -\sqrt{3}x^{3}$ corresponding to a
target drift function $U^{*}(x)$ parametrized by
$(a_{0}^{*},a_{1}^{*},a_{2}^{*}) = ({1\over 5},{9\over
10},\sqrt{{3\over 20}})$.  Figure \ref{fig:phi1}(a) shows that
starting with the initial guess $U(x)=0$, minimizing the objective
function equation (\ref{eqn:R1}) leads to accurate convergence to the
unknown target drift $U^{*}(x)$ within $\sim 10$ iterations. We find
in figure \ref{fig:phi1}(b) that a five-parameter potential is
typically a marginal case in that it can only be occasionally
reconstructed, and only after a large number of iterations. However,
we are typically not able to accurately reconstruct a potential with
six parameters (see figure \ref{fig:phi1}(c)), regardless of the
number of iterations.

\begin{figure}
\begin{center}
\includegraphics[width=2.5in]{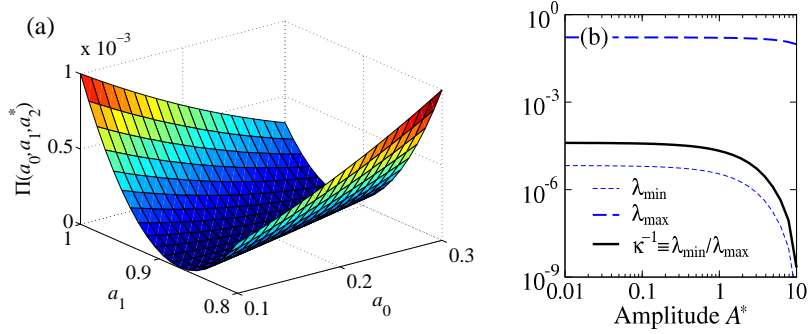}\hspace{3mm}
\includegraphics[width=1.5in]{Fig4c.eps}
\end{center}
\caption{(a) Objective function $\Pi(a_0,a_1,a_{2}^{*}) \equiv
\int_{0}^{\infty}\vert
\tilde{w}(x=0.3,s;a_0,a_1,a_{2}^{*})-\tilde{w}(0.3,s;a_{i}^{*})\vert^2\dd
s$ for the potential shown in Fig 2(a) as function of $a_0$ and $a_1$.
Projected onto $a_{0}-a_{1}$ space, $\Pi$ exhibits a much smaller
curvature in one direction compared to the orthogonal direction.  (b)
The minimum and maximum curvatures of $\Pi({\bf a})$, and the inverse
condition number. For constant weighting $g(s)$ defined in equation
(\ref{eqn:R1}), we find that the condition number $\kappa$ decreases
monotonically as the amplitude of the target potential
$A^{*}\rightarrow 0$.}
\label{fig:minimum}
\end{figure}

When the unknown target drift function $U^{*}(x)$ is structurally more
complex, extremely slow or nonconvergence to ${\bf a}^{*}$ arises
because the curvature of $\Pi({\bf a})$ near the true minimum, in at
least one direction, becomes extremely small.  Since our minimization
algorithm relies on essentially inverting the $n\times n$ Hessian
matrix $H_{ij} = \partial_{a_{i}}\partial_{a_{j}}\Pi \vert_{{\bf a}=
{\bf a}^{*}}$ (see \ref{appA}), the mathematical feasibility of the
reconstruction is limited by its condition number $\kappa \equiv
\lambda_{\rm max}/\lambda_{\rm min}$.  Here, $\lambda_{\rm max}$ and
$\lambda_{\rm min}$ are the largest and smallest eigenvalues of $H$,
representing the largest and smallest curvatures of $\Pi({\bf a})$ at
${\bf a}^{*}$, along the corresponding eigendirections, respectively.
In addition to increasing the number of eigendirections, increasing
$n$ rapidly decreases, in particular, the {\it minimum} curvature
$\lambda_{\rm min}$, thereby increasing $\kappa \equiv \lambda_{\rm
max}/\lambda_{\rm min}$ and making the minimum in $\Pi({\bf a})$
harder to find. As shown in \ref{appB}, larger values of $i$ and $j$ correspond to more
rapidly oscillating basis functions that reduce the magnitude of
$H_{ij}$.  This property renders the problem badly conditioned and is
the underlying mathematical reason for the difficulty of extracting
more than three parameters from a given potential landscape. To
explicitly illustrate the ill-conditioning of the problem, we plot in
figure \ref{fig:minimum}(a) the three-parameter objective function
$\Pi(a_{0},a_{1}, a_{2}^{*})$ (with $g(s)=1$) as a function of the
parameters $a_{0}$ and $a_{1}$. Although the global minimum
in $\Pi({\bf a})$ occurs at $(a_{0}^{*}, a_{1}^{*})$, it is clear that
the curvature near the minimum is extremely small along at least one
direction, making the minimum difficult to find numerically.

Since all three target potentials considered in figure \ref{fig:phi1}
were chosen to have $\vert {\bf a}^{*}\vert^{2} = 1$,
figure \ref{fig:phi1} fairly compares the reconstruction of
different-shaped potential functions with equal amplitude While
increasing the dimension $n$ makes the problem more ill-conditioned,
for fixed $n$, reconstruction efficiency may nonetheless depend on the
typical magnitude of the potential to be reconstructed. To compare
reconstructions of potentials of different expected magnitudes, we
define the amplitude factor

\begin{equation}
A^{*} \equiv \sqrt{\sum_{i=0}^{n-1}(a_{i}^{*})^{2}}=\vert {\bf a}^{*}\vert
\end{equation}
for each target drift function. While the amplitude $A^{*}$ needs to
be found from reconstructing the values of $a_{i}^{*}$, its value sets
the scale of the unknown drift function relative to thermal diffusion
and defines an effective Peclet number for this
problem. Experimentally, the shapes of potentials are fixed by
molecular details; however, the Peclet number $A^{*}$ is inversely
proportional to temperature and can in principle be tuned
experimentally.

Figure \ref{fig:minimum}(b) shows that for a fixed-shape target drift
function $U^{*}(x)$ of the form ${\bf a}^{*} = A^{*}\times
(1/5,9/10,\sqrt{3/20})$, and a single a FPTD measurement, the inverse
of the condition number is maximized in the limit $A^{*}\rightarrow
0$. Therefore, the problem has the best conditioning and the most
efficient mathematical reconstruction in the zero Peclet number limit,
when the potentials are weak.  Computationally, a single FPTD data set
arising from a vanishingly small drift perturbing the purely diffusive
problem gives the most numerical ``signal'' for reconstructing the
coefficients of $U^{*}(x)$.  Although the magnitudes of $a_{i}$ are
vanishingly small, their incremental effect on reducing the condition
number $\kappa$ is nonetheless greatest in this limit. This optimal
limit arises from a mathematical analysis and is not predicted by
physical considerations. However, system and experimental constraints
may preclude measurement of effective potentials at extremely low
Peclet numbers $A^{*}$ (high temperatures), suggesting that an
optimal, intermediate temperature may still arise in practice.

%
%
\begin{figure}
\begin{center}
\includegraphics[width=4.75in]{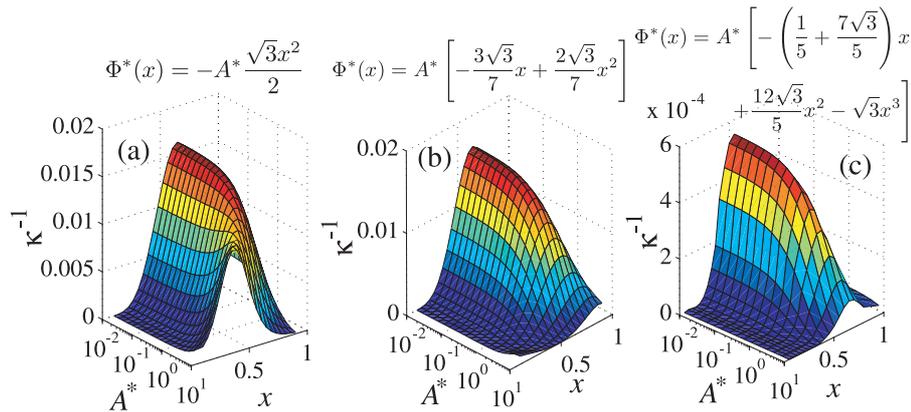}
\end{center}
\caption{Inverse condition number as a function of starting
position $x$ and drift amplitude (Peclet number) $A^{*}$ for three
different potentials.  (a) $\Phi^{*}(x) = -A^{*}{\sqrt{3}\over
2}x^2$ (b) $\Phi^{*}(x) = A^{*}\left[-{3\sqrt{3}\over 7} x +
{2\sqrt{3}\over 7}x^2\right]$ (c) $\Phi^{*}(x) =
A^{*}\left[-({1+7\sqrt{3}\over 5})x + {12\sqrt{3}\over 5}x^2
-\sqrt{3}x^3\right]$. The ratio $\kappa^{-1} = \lambda_{\rm
min}/\lambda_{\rm max}$ is typically maximal for $A^{*} \rightarrow 0$
and $x\sim 0.75$.}
\label{XD}
\end{figure}

In figure \ref{XD}, the behavior of $\kappa^{-1}$ as a function of starting position
is even more intriguing. For constant $g(s)$ and all
potentials we tested, the optimal value of the starting position
occurs roughly near $x=0.7-0.9$.  This starting position is close to
the rupture point at $x=1$ and is somewhat insensitive to the
amplitude $A^{*}$, except for very large $A^{*}$. The robustness of
this optimal starting position arises from analysing the Hessian
matrix, in particular the dependence of its condition number
on $x$.

%
%
From the form of $H_{ij}$ (see equation (\ref{eqn:S11})), we can show
numerically that $\lambda_{\rm min}$ has a maximum for $x\approx
0.7-0.9$ and that the behavior of $\kappa^{-1}$ is rather insensitive
to changes in $\lambda_{\rm max}$; thus, $\kappa^{-1}$ is typically
maximal near $x=0.75$.  For the three qualitatively different
potentials used in figure \ref{XD}, the optimal starting positions all
fall approximately within $x = 0.7-0.9$ for a wide range of amplitudes
$A^{*}$.  In these examples, the best conditioning occurs in the limit
$A^{*}\rightarrow 0$, consistent with figure \ref{fig:minimum}(b).
For nonconstant $g(s)$, the approximate optimal starting position $x$
typically ranges from 0.5 to 0.9, depending on the form of $g(s)$
(cf. figure \ref{fig:HH} in \ref{appB}).

\subsection{Multiple measurements}
\label{multiple}
Since a single measurement $\tilde{W}(s)$ is typically insufficient
to reconstruct the potential well beyond three coefficients, even
after optimizations with respect to Peclet number $A^{*}$ and starting
position $x$, we consider how additional data can be used to refine
the reconstruction of $U^{*}(x)$.  As indirectly suggested by the
analysis of varying $A^{*}$ and $x$, an unknown potential can be
changed by a specified amount to yield a FPTD different from that of
the original unchanged potential. By imposing any number of
perturbations, multiple FPTD data $\tilde{W}$ can be measured and used
to aid the reconstruction of the original potential.

We propose three protocols for modifying the potential to be
reconstructed. Experimentally, these protocols correspond to changing
the system temperature, applying then quickly removing a force to
change the starting position, and adding an applied force at the start
of the stochastic process.  Mathematically, these perturbations
correspond to specific changes in $A^{*}$, $x$, and the form of the
potential $\Phi^{*}$, respectively.  The multiple FPT distributions,
measured under different conditions, can then be combined into a
multi-distribution objective function.  We summarize the protocols
below:

\vspace{3mm}

\noindent $\bullet$ {\bf Changing amplitude via temperature} - One way
to obtain additional data is by changing the amplitude (or effective
Peclet number) $A^{*}$ of the unknown drift. For each distinct value
of $A^{*}$, a separate FPTD $\tilde{w}(x,s;{\bf a}^{*})$ can be
measured.  These different FPTD's all arise from bond potentials with
the same underlying shape, and can be used together to better 
reconstruct $\Phi^{*}(x)$.
%
%
While the absolute value of $A^{*}$ needs to be determined from the
reconstruction of ${\bf a}^{*}$, the relative temperature at which a
second measurement is taken can be used to determine the ratio
$\theta_{2}^{*}\equiv A_{2}^{*}/A_{1}^{*}$.


\vspace{3mm}

\noindent $\bullet$ {\bf Tuning starting positions} - By adding a
force to the system {\it before} the start of the process, one can
adjust the initial position $x$ of the bond.  At $t=0$, this force is
released, and the stochastic process proceeds under the original
target drift $U^{*}(x)$, provided the potential relaxes quickly to
$\Phi^{*}(x)$.  Stochastic bond dynamics starting at different
positions $x$ yield different measured FPT distributions.

\vspace{3mm}

\noindent $\bullet$ {\bf Adding Probe Forces} - Finally, one can add
known potentials to the original target potential immediately {\it
after} the start of the stochastic process to obtain additional FPT
distribution data. Here, $\Phi^{*}(x) \rightarrow \Phi^{*}(x) +
\Delta\Phi(x)$, where $\Delta\Phi(x)$ is known.  The associated drift
then changes according to $U^{*}(x) \rightarrow U^{*}(x) + \Delta
U(x)$, where $\Delta U(x)$ is implemented through a known change in
the expansion coefficients $\Delta {\bf a}$ and represents an
externally applied force imposed by {\it e.g.}, a pulling device such
as an AFM tip or an optical tweezer. The associated external potential
in such cases may be of the form $\Delta \Phi(x) = -F_{\rm ext}x
-Kx^{2}/2$, where $F_{\rm ext}$ is the externally applied
time-independent force and $K$ is the elastic response of the pulling
device. In this case, the new total bond potential
$\Phi^{*}+\Delta\Phi(x)$ induces a drift $U^{*}(x) + \Delta U$ defined
by ${\bf a}^{*} + \Delta {\bf a}$ where $\Delta a_{0} = F_{\rm
ext}+K/2$ and $\Delta a_{1} = -K/(2\sqrt{3})$. This new drift gives
rise to another, different FPTD.

\vspace{3mm}

An objective function that incorporates all $M$ FPT distributions
measured under $M$ different conditions described above can be defined
as

\begin{equation}
\Pi_{M}({\bf a}) = \sum_{m=1}^{M}\int_{0}^{\infty}[\tilde{w}(x_{m}, s;
\theta_{m}^{*}, {\bf a}+\Delta{\bf a}_{m}) -
\tilde{W}(s)]^{2}g_{m}(s) \dd s,
\label{OFM}
\end{equation}
where $x_{m}$, $\Delta {\bf a}_{m}$, and $\theta_{m}^{*}\equiv
A^{*}_{m}/A_{1}^{*}$ denote the known starting position, added pulling
force, and relative temperature of the $m^{\rm th}$ measurement,
respectively.  The function $g_{m}(s)$ weights each of the $m$
measurements differently. If the data $\tilde{W}$ in equation
(\ref{OFM}) were generated from a target drift $U^{*}(x)$, as is the
case in our analyses, then it is defined as $\tilde{W}(s) \equiv
\tilde{w}(x_{m}, s; \theta_{m}^{*}, {\bf a}^{*}+\Delta{\bf a}_{m})$.
Measured data $\tilde{W}$ should be obtained with different starting
position $x_{m}$, applied force $\Delta {\bf a}$, and/or different
temperature ratios $\theta_{m}^{*}$ with at least one of the known
parameters $x_{m}, \Delta{\bf a}_{m}, A^{*}_{m}$ different among the
$M$ measurements. Multiple data sets provide additional constraints,
increasing the curvature of the objective function near ${\bf a}^{*}$.
In general, the smallest eigenvalue $\lambda_{\rm min}$ of the Hessian
matrix associated with $\Pi_{M}$ increases with $M$.  Upon minimizing
the multi-FPTD objective function $\Pi_{M}$, we obtain ${\bf a}^{*}$.

\begin{figure}
\begin{center}
\includegraphics[width=4.75in]{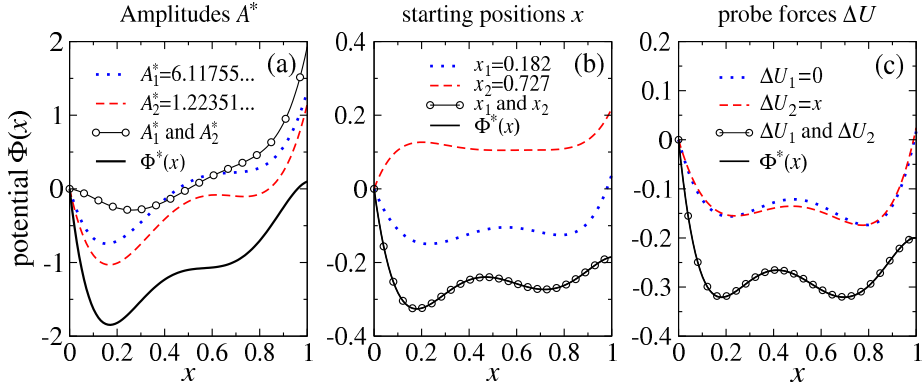}
\end{center}
\caption{ Attempted reconstruction of 5-parameter potential wells
(solid black) using single and double data sets.  (a) Blue dotted and
red dashed lines: reconstructions from single data sets generated
using $A^{*}_{1}=6.11755\ldots$ and $A^*_2 = A^*_1/5$
respectively. Circles: reconstruction using both data sets. Original
potential was parametrized by ${\bf a}^{*}=(-1/10, 11\sqrt{3}/5,
89/(14\sqrt{5}), 8/\sqrt{7}, 50/21)$, starting position was $x=0.170$.
%
%
%
(b) Blue dotted and red dashed lines: reconstructions from single data
sets generated using $x_1 = 0.182$ and $x_2 = 0.727$
respectively. Circles: reconstruction using both data sets. Original
potential was parametrized by ${\bf a}^{*}=(2/3, 2, 5/3, 2,
4/3)/\sqrt{13})$, starting position was $x=0.170$.
%
%
(c) Blue dotted and red dashed lines: reconstructions from single data
sets generated without and with probe force $\Delta U = x$.  Circles:
reconstruction using both data sets $\Delta U=0$ and $\Delta U=x$.
Target potential was parametrized by ${\bf a}^{*}=(2/3, 2, 5/3, 2,
4/3)/\sqrt{13})$, starting position was $x=0.642$.}
%
%
%
%
%
\label{fig:twoDs}
\end{figure}

To illustrate how additional data can improve potential
reconstruction, we compare how including two FPT distributions ($M=2$)
in our objective function $\Pi$ simultaneously affects reconstruction
relative to using each FPTD separately.  The two distributions will
arise from two ideal measurements taken under two different conditions
within each of the proposed experimental protocols described above.
In Figure \ref{fig:twoDs}(a), we attempt to reconstruct the
five-parameter, double-well potential $\Phi^{*}(x)=
-28x+\frac{1451}{10}x^2-307x^3+290x^4-100x^5$ using two
``temperatures.'' This potential corresponds to $A^{*}_{1}=\vert {\bf
a}^{*}\vert = 6.11755\ldots$. We then generated data associated with
$\Phi^{*}(x)/5$, corresponding to $A_{2}^{*} = A_{1}^{*}/5 =
1.22351\ldots$. Reconstruction of the original $\Phi^{*}(x)$ using
each individual FPTD fails, as does using both FPTD data sets. In
figure \ref{fig:twoDs}(b), we see that while using data sets
corresponding to either initial position $x_{1}=0.182$ or
$x_{2}=0.727$, the algorithm fails to reconstruct the target
potential. However, using both initial positions together allows us to
accurately determine $\Phi^{*}(x)$. Similarly, adding the perturbing
potential $\Delta \Phi_{2} = -x^{2}/2$ ($\Delta U_{2} = x$) provides
another FPTD that allows accurate reconstruction of a double-well
potential (figure \ref{fig:twoDs}(c)).


\section{Summary and Conclusions}

We have analysed the mathematical aspects of reconstructing the drift
of a stochastic process from perfectly measured first passage time
distributions.  In practice, insufficient number of bond rupture
events are currently measured to enable quantitative potential
reconstruction; therefore, we used numerically generated data to
illustrate our main mathematical results.
For single distributions, only very coarse attributes (approximately
three parameters) can be reconstructed. We demonstrate how to optimize
the efficiency of the reconstruction by controlling the effective
amplitude or Peclet number $A^{*}$ and starting position $x$ of the
stochastic process.  If only one FPTD can be measured, our analysis
suggests that $A^{*} \rightarrow 0$ and $x\sim 0.75$ are the most
likely parameters to give the best chance for reconstructing
relatively simple potentials.  However, these findings were found
numerically by assuming {\it perfect} data, uniform diffusivity,
precisely defined starting positions $x$, and a constant weighting
function $g(s)$.  In practice, finite time resolution, noisy data, and
other experimental limitations may be accounted for by a more suitable
weighting function $g(s)$ (or $g(t)$ if $\Pi$ were a functional of
$w(x,t)$ and the data $W(x,t)$). We show in \ref{appB} that the
optimal parameters $A^{*}$ and $x$ can change when a nonconstant
weighting function $g(s)$ is used.  While the optimal values of
$A^{*}\rightarrow 0$ and $x \sim 0.75$ (for $g(s) =1$) were found
numerically, without physically realistic limitations, they
nonetheless provide a possible experimental starting point.

%
%

We also showed that additional measurements in the form of multiple
FPTDs can be used to provide dramatically better conditioning of the
problem, allowing finer details of the drift function to be
extracted. The total objective function including the constraints from all $M$
measurements has a sharper minimum, increasing the efficiency of
standard optimization algorithms.
%
%
We proposed three ways of obtaining additional measurements under
different experimental conditions: tuning the effective amplitude or
Peclet number $A^{*}$ through the system temperature, adjusting the
starting position $x$ via an initially applied force, and adding a
known ``probe'' force through a potential $\Delta\Phi$.  This later
potential can be realized in a number of ways, from directly
mechanically pulling on the bond, to using mutagenesis to
systematically change local properties of the bond energy
profile. Such mutant bonds may provide additional FPTD data
facilitating reconstruction of the original ``wild-type'' potential.

A number of refinements are suggested by our analysis, and some are
discussed in the Appendices. For example, rather than
Laplace-transforming the data, one can directly fit to
$W(x,t)$. Although this approach is computationally more expensive, it
would allow us to treat time-dependent potentials $U(x,t)$, and
directly analyze dynamic force spectroscopy experiments (Evans
\textit{et al.} 1995, Heymann \& Grubm\"{u}ller 2000, Fuhrmann
\textit{et al.} 2008), or scenarios in which the temperature is
changed in a time-dependent way (Getfert \& Reimann 2009).  Moreover,
specific to the bond rupture problem, data involving bond coordinates
as a function of time, if accurately measured, can also be incorporated
into the objective function. These additional data may be useful in
combating the noise problem and help improve the overall conditioning.

Other more mathematical refinements and extensions can also be
implemented, including exploring effects of using different basis
functions for $U(x)$, using more sophisticated optimization methods,
quantifying the reconstruction efficiency for large $M$, defining the
experimentally-imposed weighting function $g(s)$, reconstruction of
the diffusivity $D(x)$ itself, and systematically exploring the
effects of noise in the data.  Here, a Bayesian approach to estimate
likelihood functions for the coefficients ${\bf a}^{*}$, or
information criteria to choose the size $n$ of the basis expansion
might be useful (Getfert \textit{et al.} 2009).


\begin{acknowledgements}
The authors thank S. Getfert, A. Fuhrmann, and A. Landsman for helpful
comments.  This work was supported by NSF grant DMS-0349195 and NIH
grant K25 AI058672
\end{acknowledgements}

\vspace{1cm}

\appendix{Numerical Methods}
\label{appA}

\subsection{Numerical scheme for solution of the Backward Equation}

In the forward problem, with $D(x)$ and $\Phi(x)$ given, equation
(\ref{PDE1}) can be solved using standard finite difference
schemes. Figure \ref{fig:surface}(b) shows numerically computed FPTDs
for two different starting positions with $D=L=1$, $U(x) = 4 -
10x+6x^2$.  The singular boundary condition $w(x=1,t) = \delta(t)$ is
treated by taking Laplace transforms in time of equation (\ref{PDE1})
and solving equation (\ref{ODE}) for all values of the
Laplace-transform variable $s$.  Moreover, the numerical solution of
equation (\ref{ODE}) for a set of values $s$ can be found much more
quickly than solving the full partial differential equation on a large
$x-t$ grid.  For $D=1$ and a given drift function $U(x)$, we use a
spectral method (Trefethen 2000) to solve the Laplace-transformed
Backward Equation (\ref{ODE}).  First, the spatial domain is mapped
from $[0,1]$ to $[-1,1]$ using a change of variable.  Then, the
function $\tilde{w}(x,s)$ is represented by $\tilde{w}(x_{i},s) \equiv
\tilde{w}_{i}$ and interpolated between the $N$ Chebyshev points
$x_{i} = \cos\left({(i-1)\pi\over N-1}\right)$ with polynomials.  The
resulting $N \times N$ system of equations for $\tilde{w}_{i}$ are

\begin{equation}
Q_{ij}\tilde{w}_{j} = \delta_{1,i}, 
\end{equation}
where the matrix $Q_{ij}$ is

\begin{equation}
\begin{array}{rcl}
Q_{1j} &=& \delta_{1j},  \\
Q_{ij} &=& 4 ({\bf D}_{N}^{2})_{ij} + 2 ({\bf U} {\bf D}_{N})_{ij} - s \delta_{ij}, 
\quad i=2,...,N-1, \quad j=1,2,...,N \\
Q_{N j} &=& ({\bf D}_{N})_{Nj}, \quad j=1,2,...,N, 
\end{array}
\label{eqn:S2}
\end{equation}
where ${\bf D}_{N}$ is the usual $N \times N$ pseudospectral
differentiation matrix (Trefethen 2000) and 
the diagonal matrix $\mathbf{U}$ is defined by
\begin{equation}
U_{ij} = U(x_{i})\delta_{ij}. 
\end{equation}
We used $N = 51$ spectral points in all of our computations within the
iterative algorithm. Fixed$-x$ slices of the numerically obtained
functions $\tilde{w}(x,s)$ are qualitatively similar to the plots
shown in the inset of figure \ref{fig:surface}(b).

\subsection{Evaluation of the objective function}

In our analysis, we generate data by numerically computing
distributions derived from a target drift function $U^{*}(x)$ (defined
by its polynomial coefficients ${\bf a}^{*}$). Since the data is generated
numerically, we use $\tilde{W}(x,s) \equiv \tilde{w}(x,s;{\bf a}^{*})$ in the
objective function and write
\begin{equation}
\Pi(\mathbf{a}) = \int_0^{\infty} [\tilde{w}(x,s;\mathbf{a}) - \tilde{w}(x,s;\mathbf{a}^*)]^2 g(s) \dd s.
\label{eqn:S4}
\end{equation}
%
Note that the $\ell^{\rm th}$ moment of the FPTD is given by 
\begin{equation}
\langle T^{\ell}\rangle = (-1)^{\ell}{\partial^{\ell} \tilde{w}(x,s)\over \partial s^{\ell}}\bigg|_{s=0}.
\end{equation}
Therefore for two FPTDs with identical first few moments, their first
few derivative are also identical. Such FPTDs can be distinguished
only their difference at larger values of $s$ (and the function $g(s)$
in equation (\ref{eqn:S4}) can be used to weight these differences
accordingly). Only information contained in the tails of the
Laplace-transformed distributions can distinguish two FPT
distributions with equal lower moments.  Therefore for the algorithm
described in \ref{appA} to be effective, a numerical approximation to
the integral in equation (\ref{eqn:S4}) must evaluate the integrand
for sufficiently large values of $s$.  This can be done by mapping
$s\in [0,\infty]$ to $\xi \in [0,1]$ through a change in variable
$s(\xi) = \xi/(1-\xi)$, and computing
\begin{equation}
\Pi(\mathbf{a}) = \int_0^1 [\tilde{w}(s(\xi);{\bf a}) - \tilde{w}(s(\xi); \mathbf{a}^{*})]^2 
g(s(\xi)) \frac{\dd \xi}{(1-\xi)^2}. 
\label{eqn:PI}
\end{equation}
In order to choose $g(s)$ such that $\Pi({\bf a})$ remains convergent,
we assume that $g(s)$ has no singularities in $s\in \left(0,\infty\right)$ and
consider the behavior of the integrand at the end points $s=0$ and
$s=\infty$ through the asymptotic expansions
%
%
\begin{equation}
\begin{array}{rcl}
\tilde{w}(s) &=& \int_0^{\infty} e^{-st} w(t)\dd t \\
& \sim & \displaystyle \sum_{n=0}^{\infty} \frac{w^{(n)}(0) n!}{s^{n+1}}, \qquad s \gg 1, 
\end{array}
\label{eqn:asymp1}
\end{equation}
and
\begin{equation}
\tilde{w}(s) \sim \sum_{n=0}^{\infty} \frac{s^n \tilde{w}^{(n)}(0)}{n!}, \qquad s \ll 1.
\label{eqn:asymp2}
\end{equation}
Since $w(t=0;{\bf a}) = w(t=0;{\bf a}^{*}) = 0$ for any two sets of
drift coefficients $\mathbf{a}$ and $\mathbf{a^*}$, the asymptotic
expansion (\ref{eqn:asymp1}) implies that $[\tilde{w}(s; {\bf a}) -
\tilde{w}(s;{\bf a}^{*})]^2 = O(s^{-4})$ as $s \to \infty$.  If the
first $k$ time derivatives of $w(t;{\bf a})$ and $w(t; {\bf a}^{*})$
match, equation (\ref{eqn:asymp1}) implies that $[\tilde{w}(s;{\bf a})
- \tilde{w}(s;{\bf a}^{*})]^2 = O(s^{-2(k+2)})$.  For a weighting
function of the form $g(s) = s^q$, the integrand in equation
(\ref{eqn:PI}) is $O((1-\xi)^{2-q})$ as $\xi \to 1$ and is integrable
at $\xi=1$ provided $q<3$.

Since $\tilde{w}(s=0;{\bf a}) = \tilde{w}(s=0;{\bf a}^{*}) = 1$, the
asymptotic expansion (\ref{eqn:asymp2}) implies that
$[\tilde{w}(s;{\bf a}) - \tilde{w}(s;{\bf a}^{*})]^2 = O(s^2)$ as $s
\to 0$. If the first $k$ $s$-derivatives agree (\textit{i.e.}  the
first $k$ moments of $w(t;{\bf a})$ and $w(t;{\bf a}^{*})$ are
identical), equation (\ref{eqn:asymp2}) implies that
$[\tilde{w}(s;{\bf a}) - \tilde{w}(s;{\bf a}^{*})]^2 = O(s^{2(k+1)})$.
If $g(s) = s^q$ then the integrand in equation (\ref{eqn:PI}) is
$O(\xi^{2+q})$ as $\xi \to 0$ and is integrable at $\xi=0$ provided
$q>-3$.

To summarize, if we take $A(s) = s^q$, then convergence of the
integral in equation (\ref{eqn:PI}) requires $-3 < q < 3$.  When
evaluating $\Pi({\bf a})$, a fourth order open trapezoid rule (that
does not require evaluation of the integrand at the end points) was
used and typically 100-1000 uniformly spaced trapezia were found to
give sufficient accuracy for the plots shown in Figs. \ref{fig:phi1}
and \ref{fig:twoDs}.

\subsection{Minimization of $\Pi({\bf a})$}

The linear system (\ref{ODE}) must be solved for many different values
of $s$ so that it can be used in the discrete approximation to the
objective function (\ref{eqn:R1}).  To minimize (\ref{eqn:R1}), we use
a safe-guarded Newton strategy:
\begin{equation}
{\bf a}(k+1) = {\bf a}(k) - \sigma {\bf G}^{-1}({\bf a}(k)) \nabla_{\bf a}
\Pi({\bf a}(k)).
\end{equation}
Here, ${\bf G}$ is a positive definite matrix and $\sigma>0$ is the step
size chosen to minimize $\Pi({\bf a}(k+1))$ along the descent direction 
${\bf G}^{-1}\nabla_{\bf a} \Pi$.

To compute ${\bf G}$, we adopt the following procedure. If the Hessian
$H_{ij} \equiv \partial_{a_{i}}\partial_{a_{j}} \Pi({\bf a})$ is
positive definite, we set ${\bf G} \equiv {\bf H}$. If the Hessian
${\bf H}$ is not positive definite, we choose a small Tikhonov
regularization (Vogel 2002) parameter $\alpha$ such that ${\bf H} +
\alpha {\bf I}$ is safely positive definite (${\bf I}$ is the identity
matrix) and set ${\bf G} = {\bf H}+\alpha {\bf I}$. In either case,
${\bf G}^{-1}$ is also positive definite and moving in the direction
of $-\sigma {\bf G}^{-1} \nabla_a \Pi$ is guaranteed to decrease
$\Pi({\bf a}(k+1))$ for sufficiently small $\sigma$.
The value of $\sigma$ is found by performing an exact line search to
minimize $\Pi({\bf a}(k+1))$ along the descent direction.  All
Jacobian and Hessian matrices are approximated numerically using a
suitably small $\delta{\bf a}$, typically on the order of $10^{-5}$.
The algorithm terminates when the relative change in the objective
function is less than $10^{-3}$.

We represent the drift function $U(x)$ using orthonormal polynomials
on $[0,1]$ given in equation (\ref{UBASIS}). Often, the potential and
drift arising from molecular interactions diverge as $x\rightarrow 0$
(such as in the Lennard-Jones potential).  In this case, $\Phi(x)$ and
$U(x)$ could be represented using basis functions with the correct
divergent behavior for $x \to 0$. Although the Laplace-transformed
backward equation (\ref{ODE}) has an irregular singular point in this
case, the spectral method retains its ability to find solutions as
long as $x=1$ is removed from Chebyshev grid.


\appendix{Analysis of Eigenvalues and Condition Numbers}
\label{appB}

Since the ease of minimizing $\Pi({\bf a})$ is quantified by the
condition number $\kappa \equiv
\lambda_{\textrm{max}}/\lambda_{\textrm{min}}$ of the Hessian ${\bf
H}$, we now consider the behavior of the minimum and maximum
eigenvalues $\lambda_{\rm min}$ and $\lambda_{\rm max}$.  If $\kappa =
O(1)$, the problem is well-conditioned; if $\kappa \gg 1$, the problem
is badly conditioned. Clearly, the one parameter optimization problem
has $\kappa\equiv1$ and is well-conditioned. However, when the number
of parameters increases, it is desirable to find conditions under
which $\kappa^{-1}$ is maximized.

\begin{figure}[htb]
\begin{center}
\includegraphics[width=4.6in]{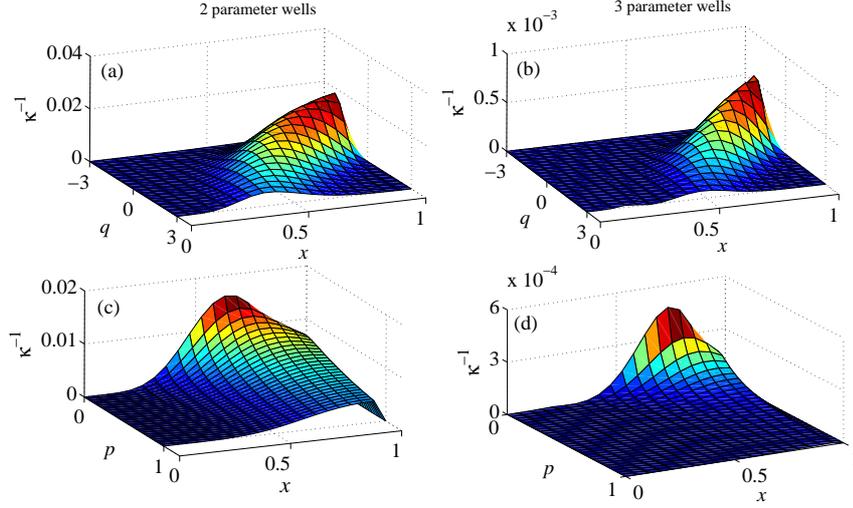}
\end{center}
\caption{Inverse condition number $\kappa^{-1}$ in the $A^{*}
\rightarrow 0$ limit as a function of starting position $x$ and the
weighting function $g(s)$ used in the objective function $\Pi({\bf
a})$.  In this limit all potentials (with a fixed number of
parameters) have the same Hessian. (a) Two parameter potentials with
weighting function $g(s) = s^q$. (b) Three parameter potentials with
weighting function $g(s) = s^q$.  (c) Two parameter potentials with
weighting function $g(s) = e^{-ps}$. (d) Three parameter potentials
with weighting function $g(s) = e^{-ps}$.}
\label{fig:HH}
\end{figure}

\subsection{Small Peclet number analysis}

We now analyze $\kappa$ in the $A^{*} \ll 1$ limit.  The components
of the Hessian at $\mathbf{a}^*$ (the drift coefficients of the target
potential) are given by
\begin{equation}
H_{ij} = \left. \frac{ \partial^2 \Pi({\bf a}) }{ \partial a_i
\partial a_j}\right|_{\mathbf{a} = \mathbf{a}^*} = 2\int_0^{\infty}
\left. g(s) \left( \pt{\tilde{w}}{a_i} \pt{\tilde{w}}{a_j}\right)\right|_{\mathbf{a} =
\mathbf{a}^*} \dd s.
\label{eqn:hessian}
%
\end{equation}  
For weak potentials, we find solutions to $\tilde{w}''
+ U(x) \tilde{w}' = s \tilde{w}$ in the $U(x) \rightarrow 0$ limit.
If $U(x) = O(A^*)$, we expand
the solution in the form $\tilde{w}(x,s) = \sum_{m=0}^{\infty} \tilde{w}_m(x,s)$
where $\tilde{w}_{m} = O({A^*}^m)$, and use the boundary conditions
$\tilde{w}_0(1,s) = 1$, $\tilde{w}'_0(0,s) = 0$ and
$\tilde{w}_m(1,s)=\tilde{w}'_m(0,s)=0$ ($m \geq 1$)
where primes denote differentiation with respect to $x$.
The first two terms
in the expansion are
\begin{eqnarray*}
\tilde{w}_0(x,s) &=& \frac{\cosh \sqrt{s} x}{\cosh \sqrt{s}}, \\
\tilde{w}_1(x,s) &=& \frac{\sqrt{s}}{\cosh \sqrt{s}} \int_0^1 G(x,x') U(x') \sinh 
\sqrt{s}x' \dd x', 
\end{eqnarray*}
where the Green's function
\begin{equation*}
G(x,x') = {\sinh\sqrt{s}(1-x_{>})\cosh\sqrt{s}x_{<} \over
\sqrt{s}\cosh\sqrt{s}},
\end{equation*}
satisfies $G'' - sG = -\delta(x-x')$ and $x_{<}(x_{>})$ is the lesser(greater) value of $x,x'$.  Upon using
the full orthonormal polynomial expansion $U(x) = \sum_{i=0}^{\infty}
a_i u_i(x)$, we find explicitly

\begin{equation}
H_{ij} =\int_0^{\infty}\!\frac{2g(s)s}{\cosh^2 \sqrt{s}} ~\dd s \int_0^1\! \int_0^1 
\!G(x,y) G(x,y') u_i(y) u_j(y') \sinh (\sqrt{s} y) \sinh (\sqrt{s} y') \dd y \dd y'. 
\label{eqn:S11}
\end{equation}
Since all elements of ${\bf H}$ are independent of the drift
coefficients $a_{i}$, they are also independent of $A^{*}$, and so are
all eigenvalues.  Therefore, as a function of $A^{*}$, the inverse
condition number $1/\kappa \equiv \lambda_{\textrm{min}}/
\lambda_{\textrm{max}}$ approaches a constant as $A^{*} \rightarrow
0$. Moreover, for all forms of $g(s)$ tested, we find numerically that
$\kappa^{-1}$ is maximal in the $A^{*} \rightarrow 0$ limit. These
results confirm the numerical data in figure \ref{fig:minimum}(b) and figure \ref{XD}.


Within the $A^{*} \rightarrow 0$ limit, different weightings $g(s)$
can also be used to better maximize $\kappa^{-1}$.  Using the
asymptotic form (\ref{eqn:S11}) for the Hessian and $g(s) = s^q$ ($-3 < q < 3$),
we plot $\kappa^{-1}$ as a function of starting position $x$ and
exponent $q$ in Fig \ref{fig:HH}. Note that $q>0$ puts more weight into the
tails of $[\tilde{w}(x,s;\mathbf{a}) -
\tilde{w}(x,s;\mathbf{a}^*)]^2$, accentuating differences in higher
moments of the FPTD.  If $q<0$, more weight is given to small values
of $s$. For each value of $q$, there is an optimal starting position
$x^*$ that minimizes the condition number and greatly improves the
efficiency of reconstructing the bond potential. Although in \S4
we discussed changing the starting positions to optimize the
reconstruction, in cases where the starting position cannot be
controlled, another strategy may be to estimate an optimal $q=q^*$
from figure \ref{fig:HH} and minimize $\Pi({\bf a})$ using the weighting function
$g(s) = s^{q^{*}}$.  We also experimented with weighting functions of
the form $g(s) = e^{-ps}$ for $p>0$: see figure \ref{fig:HH}(c) and 
\ref{fig:HH}(d).  This
class of weighting functions seems to give 
poorer conditioning compared to $g(s)=s^q$ since $\kappa^{-1}$ is generally smaller.
%
%

\subsection{Conditioning with multiple data sets}

\begin{figure}[htb]
\begin{center}
\includegraphics[width=2.75in]{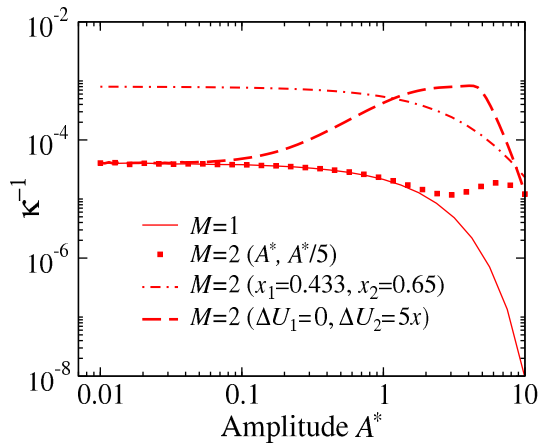} 
\end{center}
\caption{The inverse condition number $\kappa^{-1}$ as a function of
the potential's magnitude $A^*$ for one and two data sets.  The thin
solid curve corresponds to one FPTD ($M=1$), while broken curves
correspond two FPTDs ($M=2$) obtained under different conditions.  The
dotted curve represents $\kappa^{-1}$ when $A^{*}$ and $A^{*}/5$ are
used in the multi-distribution objective function $\Pi_{M}({\bf a})$
(see equation (\ref{OFM})).
The dot-dashed curve corresponds to $\kappa^{-1}$ when $x_{1}=0.433$
and $x_{2}=0.65$. Finally, the long dashed curve represents the
inverse condition number when two target drifts $U^{*}(x)$ and
$U^{*}(x) + 5x$ are used. The weighting functions $g_m(s)=1$ in all
computations.}
\label{fig:S2}
\end{figure}

In figure \ref{fig:S2}, we assumed $g(s)=1$ and plot the inverse
condition number $\kappa^{-1}= \lambda_{\rm min}/\lambda_{\rm max}$ as
a function of the Peclet number $A^{*}$.  The potential used was
proportional to the one in figure \ref{fig:phi1}(a): $\Phi^{*}(x) =
A^{*}\left[-({1+7\sqrt{3}\over 5})x + {12\sqrt{3}\over 5}x^{2}
-\sqrt{3}x^{3}\right]$. The starting position was $x=0.433$.
The thin solid curve in figure \ref{fig:S2} is taken from figure
\ref{fig:minimum}(b) and corresponds to $\kappa^{-1}$ when only one
FPTD data set ($M=1$) is used.  We compare these values to those when
two FPTDs ($M=2$) are used as data in $\Pi_{M}({\bf a})$.  The second
data set was generated using each of the protocols discussed in
Section 4(b).  The dotted, dashed-dotted and dashed curves were
generated from the Hessian of $\Pi_{M}({\bf a})$ by using both $A^{*}$
and $A^{*}/5$, two starting positions $x_{1}=0.433$ and $x_{2}=0.65$,
and both the original potential $\Phi^{*}(x)$ and
$\Phi^{*}(x)-5x^{2}/2$, respectively. In each case we see a different
improvement in the conditioning at each value of $A^{*}$.

Since the largest eigenvalue $\lambda_{\textrm{max}}$ does not change
much with increasing $M$, the improvement in conditioning is due
mostly to increasing the smallest eigenvalue $\lambda_{\textrm{min}}$
(not shown).  Reducing the Peclet number $A^{*}$ only improves the
conditioning for large $A^{*}$.  Increasing the contraction factor in
the two values of $A^{*}$ (\textit{e.g.} from $A^{*} \to A^{*}/5$ to
$A^{*} \to A^{*}/10$) further improves the conditioning by increasing
$\kappa^{-1}$ at ever smaller values of $A^{*}$.  In this example,
changing the starting position is perhaps the most reliable way of
facilitating the reconstruction: the condition numbers $\kappa$ are
decreased by at least an order of magnitude over a wide range of
$A^{*}$ and the improvement becomes even better for {\it larger}
$A^{*}$.
Finally adding a
probe force greatly enhances the reconstruction for moderate
$A^{*}=O(1)$ and there is now an optimal $A^{*}$ (and hence system
temperature) where the reconstruction is easiest. This is in contrast
to the $M=1$ case, here we always find a monotonically increasing
$\kappa^{-1}$ as $A^{*}$ decreases.


\appendix{Reconstruction of many-parameter potentials}

Using additional FPT data, we demonstrate the feasibility of
reconstructing complex potentials that are described by many
parameters.
\begin{figure}[htb]
\begin{center}
\includegraphics[width=4.25in]{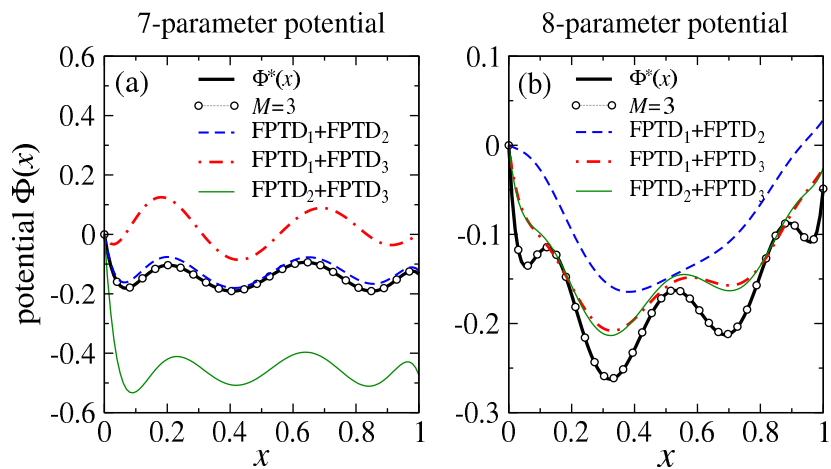} 
\end{center}
\caption{Reconstruction of seven and eight parameter potentials with
unit amplitude $A^{*}=1$ and drift coefficients (a) ${\bf a}^* =$
$(1/2, 1, 1, 3/2, 1, 3/2, 1)/\sqrt{14}$, and (b) $(1/2, 5, 2, 5/2, 1,
5/2, 2, 15/2)/\sqrt{103}$.
%
%
In (a) the three different protocols, $(x_{1}=0.07, \Delta U_{1} =
0)$, $(x_{2}=0.404, \Delta U_{2} = 0)$ and $(x_{3}=0.404, \Delta U_{3}
= x)$, were used to generate three data sets (in each case,
$A^{*}=1$), denoted by FPTD$_{1}$, FPTD$_{2}$, and FPTD$_{3}$,
respectively.
%
%
In (b) the three different protocols $(x_{1}=0.06, \Delta U_{1}=0)$,
$(x_{2}=0.323,\Delta U_{2}=0)$, and $(x_{3}=0.323, \Delta U_{3} = 5x)$
were used to generate three data sets also denoted by FPTD$_{1}$,
FPTD$_{2}$, and FPTD$_{3}$ respectively.}
\label{fig:7param}
\end{figure}

In figure \ref{fig:7param} we successfully reconstruct 7-parameter and
8-parameter potential wells containing multiple minima using $g(s)=1$
and three ($M=3$) FPTDs.  For both (a) the 7-parameter and (b) the
8-parameter potential, we see that any combination of two first
passage time distribution data sets fails to accurately reproduce the
original $\Phi^{*}(x)$. However, using the three defined data sets
($M=3$), we were able converge to the correct $\Phi^{*}(x)$ in fewer
than 20 iterations. Importantly, we were able to quantitatively
resolve the multiple minima.  It should be noted, however, that with
the current optimization algorithm we were only able to obtain about 2
significant digits of accuracy on the coefficients $a_i$ for these
potentials.

\label{lastpage}

\end{document}